# Information-theoretic model of self-organizing fullerenes and the emergence of C$_{60}$

Shoaib Ahmad

NCP, QAU Campus, Shahdara Valley, Islamabad 44000, Pakistan

Email: sahmad.ncp@gmail.com

**Abstract**

An information-theoretic model describes the dissipative dynamical systems composed of ensembles of fragmenting, self-organizing fullerenes. The probabilities derived from the variations of the fullerene number densities that occur during $cage \rightarrow cage$ transformations are used to evaluate Shannon entropies for every carbon cage. Fractal dimension of the cages are calculated from their respective entropies. C$_{60}$ is shown to emerges as the end-directed evolution of dynamical systems of four different ensembles of fullerenes. The information generating, transforming cages of carbon provide a perspective to evaluate the self-organizational behavior of dissipative structures.

Keywords: Fractal dimension, information-theoretic entropy, self-organization, fullerenes, emergence, dynamical systems

## 1. Introduction

Amongst the nanostructured allotropes of carbon [1] which include fullerenes, nanotubes and the planar sheets of graphene, fullerenes have the unique distinction of cage closure, two dimensional curvatures, isomeric diversity [2], closed cages' steric strain [3] and the ability for mutual transformation. Fullerenes demonstrate $cage \to cage$ transformations during their formative and fragmentation stages. For the optimized cage-transformation conditions, icosahedral $C_{60}$ is shown to be the sole survivor [4,5]. Self-organization of fullerenes in hot carbon vapor is treated here as dissipative dynamical systems [6-8] whose configuration changes with time. We evaluate the nonlinear interactions among the dynamical system's constituents and their interdependence. Fragmenting, re-forming fullerenes and an evolving gas of $C_2$ emerge as the constituents. Shannon entropies [9,10] of the constituents at every stage of the dynamic transitions are calculated from the iterations of the $cage \to cage + C_2$ mapping. The fractal dimension [11] can be defined by using the information-theoretic entropy as the information dimension; both definitions are equivalent as demonstrated by Renyi [12] and documented with broad range of experimental verifications in ref [13]. In this article, the transformation sequences of the entire fullerene ensembles into $C_{60}$ and $C_2$ are profiled and the information dimensions of the fragmenting and evolving fullerenes are evaluated. Our model describes the conditions for the emergence of the Buckyball-$C_{60}$. $C_{60}$'s emergence is discussed in the context of similar measures that have been discussed like the maximum entropy production [14-15], information-theoretic self-organization and emergence [16] and the end-directed evolution [17].

The driving forces of fragmentation are related to structural instability of fullerenes due to the pentagon-related internal stresses [18] and degeneracy pressures of the Fermi gas of pi electrons [19]. Hot carbon vapor in an appropriately confined environment can generate fullerenes with relative densities that are determined by their respective heats of formation and isomeric abundance. Various mechanisms and routes for the emergence of $C_{60}$ out of the disorder of the hot C-vapor have been proposed [18-27]. In an earlier, thermodynamic entropic model, we treated the forming and fragmenting fullerenes of the grand canonical ensemble as 3D rotors. The thermodynamic entropy of cage-to-cage transformations was evaluated [27]. The thermodynamic entropic arguments are re-evaluated here by Shannon entropic considerations. We discuss the advantages in employing information-theoretic entropy over thermodynamic entropic arguments in describing the self-organization of carbon cages. We show that the information-theoretic technique is not only superior to thermodynamic model but it can be extended to other similar systems. It was recently applied to the linear and nonlinear dissipative structures in irradiated single-walled carbon nanotubes [28].

The probability $p_x(\zeta)$ is calculated for the sequences of fragmentation stages $\zeta$ of the dynamic transformation of fullerenes $C_x$ containing *x* carbon atoms. Shannon entropy is the sum of the instantaneous entropy $-p_x(\zeta)lnp_x(\zeta)$ for each fullerene over all fragmenting stages. It is used to calculate fractal dimension of the constituents of the dynamical system. Fractal dimensional analysis of the self-organizing carbon cages describes their interactions, the nature of transitions and the evolution of $C_{60}$. We believe, to the best of our knowledge, that this is the first attempt of describing the emergence of the $C_{60}$ by fractal analysis of ensembles of self-organizing fullerenes.

## 2. The closed cages of carbon-fullerenes

Carbon vapor condenses with higher probabilities for clusters $C_x$ with increasing number of carbon atoms-$x$. Linear chains, rings and sheets have single or few isomers for each type. For $x \geq 20$ the closed cages start to form and thereafter, have an increasing number of isomers for each $x$. The cluster growth is dominated by large closed cages whose respective heats of formation $E_x$ reduce with the increasing $x$ while the number of isomers $F(x)$ for each fullerene has a power aw dependence. The closed cages of carbon are formed with higher probabilities and larger number of isomers. The initial bottom-up sequence of the increasingly larger cages' formation is followed by the subsequent top-down shrinking cascades. In this communication, the top-down sequence of the fragmenting, self-organizing carbon cages is investigated by information-theoretic methods [16,28].

*2.1. Structural variability*
After being formed, unlike graphene or nanotubes, the fullerenes possess the property of inter-cage structural variability. A fullerene can be formed in a number of isomeric possibilities. The fullerene with 60 carbon atoms can be formed with any of the 1812 isomeric structures that involve the statistical combinations of the 12 pentagons among 20 hexagons. For $C_{70}$ the number of isomers is 8149 and for $C_{80}$ the isomeric possibilities grow to 31,924 [2]. A power law relates the number of isomers $F(x)$ with the number of carbon atoms in a fullerene as $F(x) \sim x^n$, where $n$ can be associated with the isomeric dimension of the fullerene $C_x$. The power law-based structural variability induces an isomeric dimension that can be defined using the dimensional analysis [12,13]

$$n \equiv d_F^x = -\ln(F(x))/\ln(1/x) \tag{1}$$

Here $F(x)$ is the total number of $d_F^x$ dimensional units with $x$ cages. This definition resembles the fractal dimension of a self-similar structure. But all isomeric variations of $C_{60}$ may not necessarily be present in any ensemble of $C_{60}$s. This statistical aspect restricts the use of the derived values of $d_F^x$ as the fractal dimension for the fullerene $C_x$. The isomeric dimension $d_F^x$ of $C_{60}$, $C_{70}$, $C_{80}$, $C_{90}$ and $C_{100}$ calculated by using equation (1) are 1.83, 2.12, 2.37, 2.56 and 2.73, respectively. The power law for the isomeric diversity introduces the statistical nature of the structural variability of fullerenes. It demonstrates the ability of a fullerene to fragment to a number of available isomers of the next smaller one.

*2.2. The inter- and intra-cage rearrangements and isomer change mechanism*

The cage structure of fullerenes allows inter-cage, cage-to-cage transformations as indicated earlier. A cage in the environment of hot carbon vapor can absorb or emit $C_2$. While absorbing $C_2$s, a cage can grow until its structural instabilities force the cage's disintegration, or the $C_2$ extrusion process may initiate and eventually it may cease to exist as a cage. Once a closed cage fullerene is formed, its growth and shrinkage are determined by the ambient conditions of the formative environment in the experimental chamber. The Stone and Wales rearrangement for the motion of the twelve omnipresent pentagons have been proposed to operate on the surfaces of all cages to re-orient the pentagons and hexagons from thermodynamically unfavorable fullerene cages into more favored structures [28]. This scheme allows the intra-fullerene mechanism to operate for isomer changes. It provides a route for a non-icosahedral $C_{60}$ cage to transform into the icosahedral $C_{60}$-the Buckyball [2,29,30]. Our present analysis does not explicitly involve this mechanism. However, we have shown elsewhere that the evolution the $C_2$ gas during the fragmentation sequences, may effectively be used as an agent for intra-fullerene isomer changes [19,27].

*2.3. Fractal dimension*

The probability paradigm is at the heart of the model. It is used to define the dissipative structure, the relative performance of its constituents and the end-directed output. The normalized probability distributions are evaluated for each constituent of the dynamical system consisting of fragmenting cages. For the calculations reported here, the number of isomers are the starting number density of each fullerene. The densities of the fragmenting fullerenes are normalized with the initial isomeric density to derive the probability as a function of the successive fragmenting stages $\zeta = 0,1,2,3 \ldots$ as $p_x(\zeta)$. The associated Shannon entropy or information $I_x$ [9] is

$$I_x = -\sum_\zeta p_x(\zeta) \ln(p_x(\zeta)) \qquad (2).$$

The complete sequence of $\zeta$ steps for the cage transformation generates the net information obtained from the probability distribution $p_x(\zeta)$. Fractal or the information dimension for the inter-cage transformation is defined as [12,13]

$$d_I^x = -\sum_\zeta p_x(\zeta) \ln(p_x(\zeta))/(\ln(1/\zeta)) = I_x/\ln(1/\zeta) \qquad (3).$$

In equation (3), $1/\zeta$ is derived from total number of fragmentation stages, it acts as the measure required to obtain information for the required transformation. The calculated fractal dimension $d_I^x$ could refer to the fragmenting as well as the evolving cages. Fractal dimensional and chaos geometric analyses are powerful tools and these have been successfully employed in a number of experimental and simulation studies [31,32].

*2.4. Icosahedral $C_{60}$*

Icosahedral group $I_h$ to which only one isomer each of $C_{20}$, $C_{60}$, $C_{80}$, $C_{180}$, $C_{240}$, $C_{540}$… belong, have the highest symmetry [1,2]. The three well known fullerenes $C_{20}$, $C_{60}$, $C_{80}$ have the same icosahedral symmetry $I_h$ but in the case of $C_{20}$, there is 1 cage and only 1 isomer; for $C_{60}$, 1 out of 1812 isomers and 1 out of 31,924 for $C_{80}$. These have very different surface topographies with only $C_{60}$ having uniform distribution of the strain due to curvature of the perfect spherical shell. Surface deformations of all others lead to non-uniformity of the structural strain. High symmetry is one of the fundamental conditions with greater probabilities of the spheroidal cages' survival in hot carbon vapor [3,19]. And this can be achieved for $C_{60}$ and higher cages by fullerene isomerization through Stone-Wales rearrangement [2,29,30].

*2.5. The $C_2$ gas*

The $C_2$ gas needs a clarification. The essential characteristics are presented here; the details are discussed in ref. [25,27]. Every cage fragments into a smaller one by losing one hexagon and emitting a $C_2$ molecule. That generates, at times increasing and at other instances, variable density of the $C_2$s. On the other hand, the density of $C_{60}$s continues to increase as a stable cage. $C_2$ being a radical, diatomic molecule can be consumed in collisions with other $C_2$s to make larger molecules. Collisions with the walls of the experimental chambers also consume a major chunk of the evolving $C_2$ gas. $C_2$'s ingestion into fullerenes ($C_2 + C_x \rightarrow C_{x+2}$) can also occur as a step for the self-organizing cages [23-25]. Cooling of the hot cages may also occur via $C_2$ emission [26]. Multiple ingestions and $C_2$-spitting sequences can lead to fullerenes with higher symmetries with uniform distribution of the spherical strain. $C_2$ ingestion and extrusion mechanisms, along with the Stone-Wales processes have been identified as agents of the cages' reorganization and shrinking [24]. Simultaneously, the bond-reorienting, Stone-Wales reaction [29,30] operate and lead to icosahedral $C_{60}$. $C_2$ gas is a vanishing constituent of the fragmentation sequences, just like the larger cages that transform into the smaller ones, contribute to the information generation and fragment away.

## 3. Cage-to-cage transformations

Cage-fragmentation in a hot grand canonical ensemble of fullerenes, initiates the top-down shrinking sequences for the fullerenes $> C_{60}$. The cage $C_x$ transforms into a smaller cage and a $C_2$ molecule via reaction $C_x \rightarrow C_{x-2} + C_2$ [1,2,4]. This results in the changes of the configuration of the dynamical system composed of fullerenes of various sizes and number densities with each fragmentation step. The $C_2$-extrusion results in the decrease of fullerene size. The reverse process can also happen by $C_2$-ingestion. Together, these reactions are the primary agents of self-organization. The fragmenting process is shown as a dissipative dynamical system with mapping of an *x*-atom fullerene $C_x$ onto (*x*-2)-atom cage as

$$f: C_x \rightarrow C_{x-2} + C_2 \tag{4}$$

The iterations of cage-to-cage transformations in equation (4) have been carried out for four sets of self-organizing, fragmenting dynamical systems of fullerenes $\sum_{60}^{70} C_x$, $\sum_{60}^{80} C_x$, $\sum_{60}^{90} C_x$ and $\sum_{60}^{100} C_x$. The summation notation for ensembles of fullerenes $C_x$ implies that the ensemble contains from the smallest (lower limit) to the largest fullerene shown as the upper limit on the summation sign. For example, $\sum_{60}^{70} C_x$ consist of the array of six fullerenes $C_{60}$, $C_{62}$, $C_{64}$, $C_{66}$, $C_{68}$ and $C_{70}$.

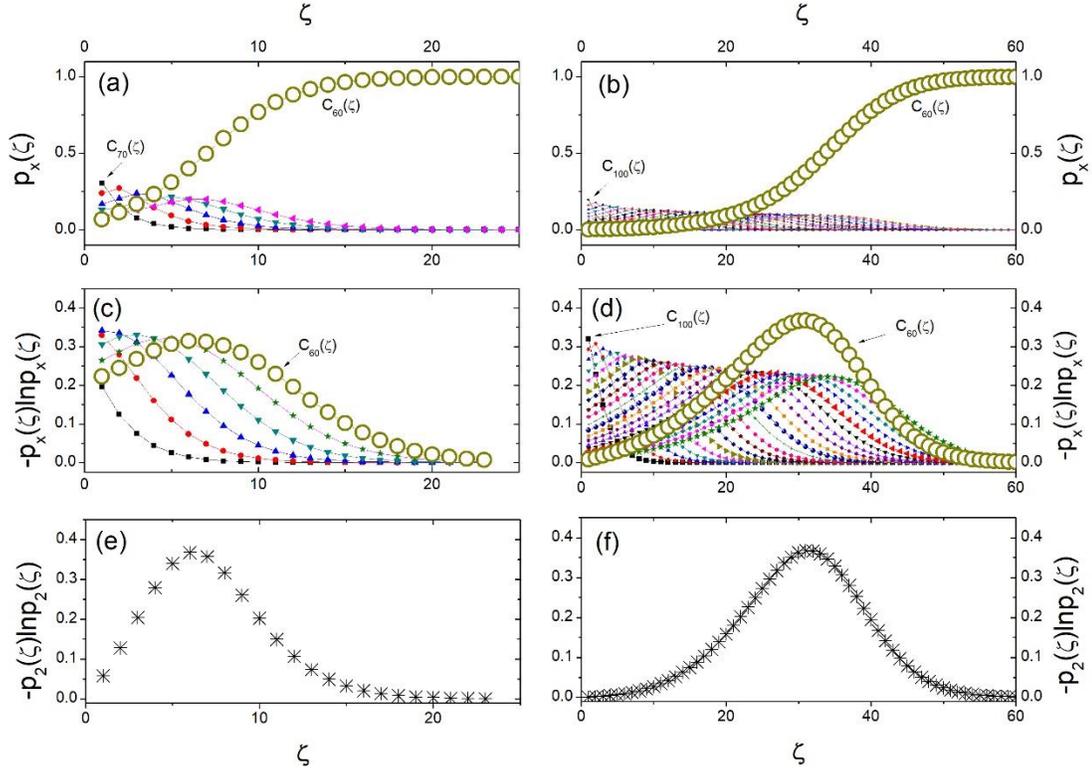

**Fig. 1.** (a) and (b) show the $Cage \rightarrow Cage + C_2$ transformations with the probabilities of fullerenes as a function of fragmentation step $\zeta$ for two sets of ensembles of cages $\sum_{60}^{70} C_x$ and $\sum_{60}^{100} C_x$, respectively. At fragmentation step $\zeta=0$, all cages start with their respective isomeric densities. In Figs. 1(c) and 1(d) the instantaneous entropy $-p_x(\zeta)lnp_x(\zeta)$ is computed and plotted at every fragmentation step $\zeta$. Gradual build-up of $C_{60}$ is visible. Fig. 1(e) and 1(f) are the corresponding entropic profile $-p_2(\zeta)lnp_2(\zeta)$ versus $\zeta$ for the gas of $C_2$s evolving in the two fragmenting ensembles.

In Figs. 1(a) and (b) the results of iterations of equation (4) of transformations of every fullerene cage except $C_{60}$, for the two sets of ensembles $\sum_{60}^{70} C_x$ and $\sum_{60}^{100} C_x$ are shown for each fragmentation step $\zeta$. The starting abundances of fullerenes are determined by their respective isomeric abundance. The data for the isomeric abundance is taken from ref. [2]. From the starting number densities, the normalized probability $p_x(\zeta)$ is evaluated for each fullerene $C_x$ of the ensembles, at successive fragmentation stages $\zeta$. It is assumed in the model that at each $\zeta$ step, half of all the cages fragment, except the $C_{60}$. Starting from $\zeta = 0$, where initial isomeric densities are assumed, $p_x(\zeta)$ is calculated for each fullerene and shown in Figs. 1(a) and 1(b). In Fig. 1(a) the probabilities of cage transformations for $C_{70}$, $C_{68}$, $C_{66}$, $C_{64}$, $C_{62}$ and $C_{60}$ are shown. In Fig. 1(b) the $p_x(\zeta)$ for the ensemble of 21 fullerenes from $C_{100}$, $C_{98}$, …… $C_{62}$ and $C_{60}$ are plotted as a function of $\zeta$. To illustrate the pattern of fragmentation, the largest fullerenes $C_{70}$ and $C_{100}$ in both the ensembles are pointed by arrows. Their respective probabilities $p_x(\zeta)$ reduce to almost zero within 5-11 fragmentation steps. All other fullerenes also follow the same route but with varying profiles of $p_x(\zeta)$ versus $\zeta$. All fullerenes shrink and fragment such that all $C_x \to C_{x-2} \to \cdots \to C_{60}$. The probability profile for $p_{60}(\zeta)$ rises from $0 \to 1$ and $C_{60}$ acts as the end-directed evolution of the dynamical system of the transforming fullerenes in Figs. 1(a) and 1(b). Figs. 1(c) and 1(d) show the entropic profile for all cage-to-cage transformations for the two chosen ensembles. The profiles for $-p_x(\zeta)lnp_x(\zeta)$ for each cage is shown where the $-p_{60}(\zeta)lnp_{60}(\zeta)$ graphs for $C_{60}$ can be seen emerging out of the two ensembles. The corresponding plots of $-p_2(\zeta)lnp_2(\zeta)$ for the evolving $C_2$ gas are shown in Fig. 1(e) and 1(f). The increasing entropy of the $C_2$ gas complements that of the $C_{60}$ cages. The emergence of $C_{60}$ in both ensembles is shown as the end-result of the sequences of cumulative fragmentations of the larger fullerenes into the next smaller ones. All fullerenes can be seen to self-organize themselves into the bell-shaped, entropic curve of $C_{60}$ as the net outcome of the dissipative, nonlinear dynamical system of the fragmenting fullerenes.

## 4. The emergence of $C_{60}$

The sum of entropies of all cages at each successive fragmentation stages $\zeta$ defines the state of the dynamical system as

$$I_\zeta = -\sum_x p_x(\zeta) \ln p_x(\zeta) \qquad (5).$$

The orbit of $I_\zeta$ defines the phase space of the self-organizing fullerenes. The four sets of spectra of $I_\zeta$ versus $\zeta$ are the phase trajectories shown in Fig. 2(a). The four spectra represent the sum of entropies of all cages of the respective ensembles at each fragmentation stage $\zeta$. Each $I_\zeta$ spectrum is the orbit of the whole of the ensemble during fragmentation. The phase trajectories of the dynamical system of the transformation of large fullerenes into the successively smaller ones represent the dynamic profile of the self-organization of fullerenes towards $C_{60}$. The dynamic pathways for transition of the entire ensemble of fullerenes into the two gases of $C_{60}$ and $C_2$ can be represented as $\sum_\zeta \sum_x C_x(\zeta) \to \sum_\zeta C_{60}(\zeta) + \sum_\zeta C_2(\zeta)$. The spectra of the phase trajectories in Fig. 2(a) represent the sum of entropies of all cages $-\sum_x p_x(\zeta) \ln p_x(\zeta)$ of the four chosen ensembles. We have included all cage transformations of the type $C_x \to C_{x-2} + C_2 \to C_{x-4} + C_2 \to \cdots C_{60} + C_2$. The sum over all $x$ and $\zeta$ of $-p_x(\zeta) \ln p_x(\zeta)$ is calculated as the entropic cost of the dynamical transition of the entire ensemble of fullerenes $\sum_x C_x(\zeta)$ into the two gases of $C_{60}$ and $C_2$. The phase trajectories of transformation of the ensembles represent the dynamic profile of the self-organizing fullerenes. The total number of cages remains constant in $cage \to cage$ transformations.

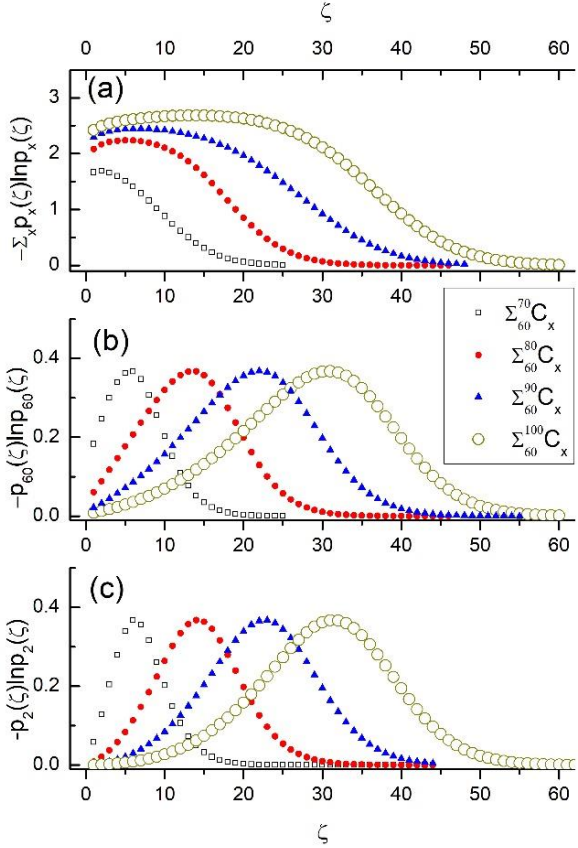

**Fig. 2.** (a) Phase space profiles of the shrinking and disappearing fullerenes. The symbols for the four dynamical systems of the fullerene ensembles are shown in the **inset**. Each point of the respective phase trajectories represents the state of all the fragmenting and evolving cages. Each state is shown as the sum $-\sum_x p_x(\zeta) \ln p_x(\zeta)$ of all the cages. (b) shows the corresponding phase trajectories of the evolving $C_{60}$ gas $-p_{60}(\zeta) \ln p_{60}(\zeta)$ as a function of $\zeta$. Similarly, the evolution of the $C_2$ gas is shown for each $\zeta$ in (c). **Inset:** shows the symbols for the four ensembles.

## 5. Fractal dimension of self-organizing cages

From the graphs of $-p_x(\zeta) \ln p_x(\zeta)$ against $\zeta$ plotted in Fig. 1, the information [9,10] or Shannon entropy for the shrinking ($>C_{60}$) and the accumulating ($C_{60}$) fullerenes is calculated from

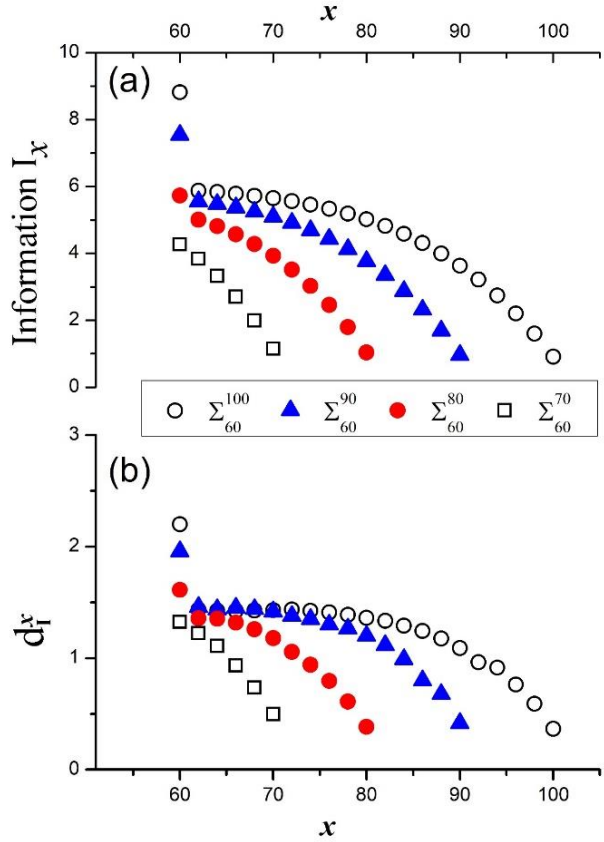

**Fig. 3.** Fractal description of self-organizing fullerenes. (a). The Shannon entropy or the Information $I_x$ from equation (2) is plotted for all transforming cages of the four dynamical system shown in the inset. Information maximizes for all the ensembles as the fragmenting cages shrink towards $C_{60}$. (b). Fractal dimensions $\boldsymbol{d_I^x}$ are evaluated using $I_x$ and plotted for every constituent fullerene by using equation (3). **Inset:** the symbols for the fullerenes belonging to the four ensembles are shown.

$I_x$ in equation (2). It is the area under the entropic profiles or the trajectories for each of the fullerenes and the associated $C_2$s. Information $I_x$ or Shannon entropy of the self-organizing fullerenes for the four ensembles $\sum_{60}^{70} C_x$, $\sum_{60}^{80} C_x$, $\sum_{60}^{90} C_x$ and $\sum_{60}^{100} C_x$ is shown in Fig. 3(a). In the model presented here, fullerenes belonging to any of the four ensembles, start with their initial, pre-fragmentation densities. Their densities are reduced by half at every fragmentation step. The largest fullerenes fragment away within 10-15 $\zeta$ stages. The smallest one, $C_{60}$, being the recipient of all of the cage-transformations, has the entire fragmenting ensemble as its basin. Therefore, it has the largest number of accumulation stages ~ the maximum number of $\zeta$. Information obtained by summing $-p_x(\zeta)lnp_x(\zeta)$ for every cage, over all $\zeta$ follows the trend $I_x < I_{x-2} < I_{x-4} ... < I_{60}$ for the fullerenes of the four ensembles. This pattern is clearly visible in Fig. 3(a). The cumulative results for the ensemble show the fragmentation pattern of fullerenes. The sequences of all cage fragmentations lead to the enrichment of a single species i.e. $C_{60}$. The largest ensemble also has the highest $I_x$ for $C_{60}$. In Fig. 3(b) the fractal dimensions $d_I^x$ of the constituent fullerenes calculated by using equation (3) are plotted. Fractal dimension varies for all fullerenes of the ensemble. It follows the trend shown by the corresponding information $I_x$ in 3(a). The dynamical system with wider range of $x$ like $\sum_{60}^{100} C_x$ yield higher fractal dimension for $C_{60}$. Fractal dimension $d_I^{60}$ for the accumulating $C_{60}$ gas increases with the size of the ensemble. This trend is a natural consequence of the availability of the increasing numbers of larger ($> C_{60}$) cages. Therefore, due to higher abundance for the larger fullerenes with lower cage formation energies, the starting ensemble with successively larger fullerenes will always have increasing number densities for cages $> C_{60}$ and hence the largest fractal dimension $d_I^{60}$ for $C_{60}$. Similarly, the fractal dimension of the accumulating $C_2$-gas can be obtained from $d_I^2 = -\sum_\zeta p_2(\zeta)lnp_2(\zeta)/ln(1/\zeta)$. Fractal dimension $d_I^x$ for each fullerene shown in Fig. 3(b) is obtained from the respective information $I_x$. $C_{60}$ has the highest fractal dimension $d_I^{60} > d_I^{x>60}$. It emerges as the end-directed evolution [17] of the dynamical system of fragmenting fullerenes.

**Table 1.** Information for all of the cages of the four ensembles, the emerging gases of $C_{60}$ and $C_2$ are tabulated. These are shown as $\sum I_x \equiv -\sum_\zeta \sum_x p_x(\zeta)\ln(p_x(\zeta))$, $I_{60} = -\sum_\zeta p_{60}(\zeta)\ln(p_{60}(\zeta))$ and $I_2 = -\sum_\zeta p_2(\zeta)\ln(p_2(\zeta))$. The fractal dimension $d_I^{60}$ of the emergent structure $C_{60}$ and the $C_2$ gas are tabulated. The ratio of entropies of the fragmenting cages to that of the emerging one as $i_{60}^\Sigma = \sum I_x / I_{60}$ is also shown, it starts to reach a saturation stage for large ensembles.

|  | $\sum_{60}^{70} C_x$ | $\sum_{60}^{80} C_x$ | $\sum_{60}^{90} C_x$ | $\sum_{60}^{100} C_x$ |
|---|---|---|---|---|
| $\sum I_x$ | 23.34 | 50.79 | 79.42 | 112.8 |
| $I_{60}$ | 3.4 | 5.73 | 7.54 | 8.83 |
| $d_I^{60}$ | 1.13 | 1.61 | 1.9 | 2.1 |
| $I_2$ | 2.66 | 4.88 | 6.39 | 7.67 |
| $d_I^2$ | 0.99 | 1.37 | 1.68 | 1.87 |
| $i_{60}^\Sigma$ | 6.86 | 8.86 | 10.53 | 12.77 |

Information is shown to be the crucial link between the phase space of a dissipative dynamical system and fractal dimensions of its constituents. The data obtained for the four dynamical systems are tabulated as Table 1. It has the cumulative Information for all of the cages as $-\sum_\zeta \sum_x p_x(\zeta) ln p_x(\zeta)$, for the evolving gas of $C_{60}$s it is calculated by $-\sum_\zeta p_{60}(\zeta) ln p_{60}(\zeta)$ and for the $C_2$s gas as $-\sum_\zeta p_2(\zeta) ln p_2(\zeta)$. The fractal dimensions $d_I^{60}$ and $d_I^2$ of the accumulating gases of $C_{60}$ and $C_2$ have been calculated and displayed for the four ensembles. The ratio of the total information of the fragmenting fullerenes and the $C_2$ gas to that of the emerging $C_{60}$s is defined as $i_{60}^\Sigma \equiv I_{total}/I_{60}$. It has numerical values 6.86, 8.86, 10.53 and 12.77 for the four ensembles $\sum_{60}^{70} C_x$, $\sum_{60}^{80} C_x$, $\sum_{60}^{90} C_x$ and $\sum_{60}^{100} C_x$. Similarly the fractal dimension $d_I^{60}$ is evaluated for the fragmenting fullerenes of the dynamical systems $\sum_\zeta \sum_x C_x(\zeta) \to \sum_\zeta C_{60}(\zeta) + \sum_\zeta C_2(\zeta)$, where the larger cages transform into C$_{60}$. It yields successively increasing the information dimension $d_I^{60}$= 1.13 to 2.1 for the 4 dynamical systems.

## 6. Information-theoretic measures of emergence: $ln(1/p_x(\zeta))$ and $-p_x(\zeta)lnp_x(\zeta)$

Our model deals with two kinds of fullerenes; ones that fragment, transform into the smaller ones and disappear, while the other kind is the emerging fullerene into whom all others transform. In addition to the information $I_x = -\sum_\zeta p_x(\zeta)lnp_x(\zeta)$, the instantaneous entropy $-p_x(\zeta)lnp_x(\zeta)$ and the function $ln(1/p_x(\zeta))$ is used here to demonstrate the information-theoretic functional dependence of the fragmenting and emerging fullerenes. For the series of the fragmenting stages $\zeta$ of a fullerene $C_x(\zeta)$ with probability $p_x(\zeta)$, the uncertainty is defined as $ln(1/p_x(\zeta))$. When it is plotted against $\zeta$, it describes the local and the fullerene-specific information. This function implies that lower the probability, the higher the 'surprise' [33]. Shannon entropy $I_x = \sum_\zeta -p_x(\zeta)lnp_x(\zeta)$ in equation (2) is the weighted average or the expected value of $ln(1/p_x(\zeta))$ which implies that maximum entropy is the least information.

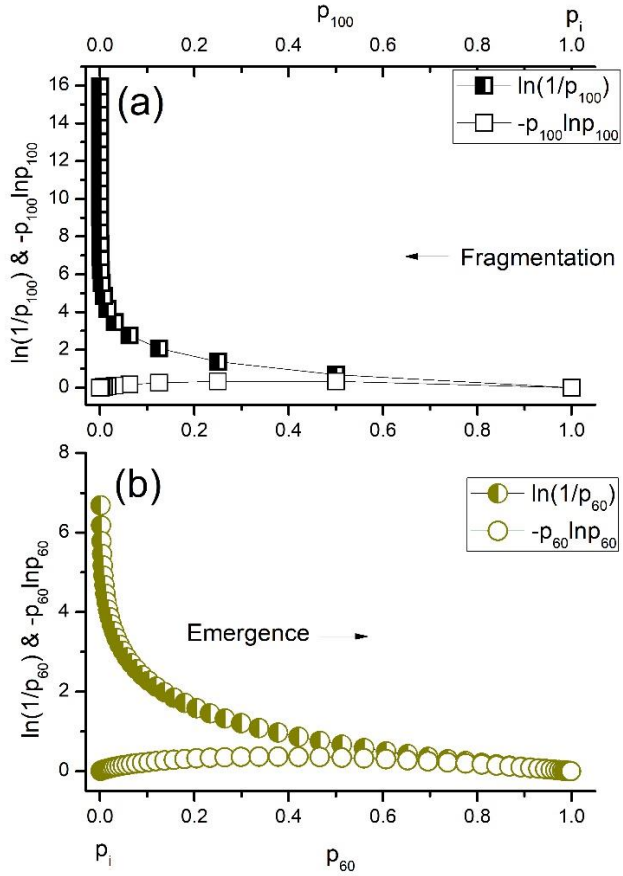

**Fig. 4.** $ln(1/p_x(\zeta))$ and $-p_x(\zeta)lnp_x(\zeta)$ are plotted as a function of the respective probability $p_x(\zeta)$ in (a) for the fragmentation of $C_{100}$ and (b) for the evolution of $C_{60}$. These results are for the ensemble $\sum_{60}^{100} C_x$. Maximum 'surprise' is near low probability and minimum around $p_x(\zeta)\sim 1$. The initial probabilities $p_i$ for the two are shown. The directions are indicated by arrows for $C_{100}$'s fragmentation and the emergence of $C_{60}$.

The information-theoretic functions, $ln(1/p_x(\zeta))$ and $-p_x(\zeta)lnp_x(\zeta)$, are plotted against $p_x(\zeta)$ in Fig. 4 (a) for $C_{100}$ and in (b) for $C_{60}$ for the ensemble of fragmenting fullerenes $\sum_{60}^{100} C_x$. When plotted together, $ln(1/p_x(\zeta))$ and $-p_x(\zeta)lnp_x(\zeta)$ describe the fragmenting and disappearing fullerenes with the emerging, self-organizing fullerenes. In 4(a) the entire population of $C_{100}$ fragments away as $p_x(\zeta) \to 0$. That is where the maximum 'surprise' occurs as demonstrated by $ln(1/p_x(\zeta))$ versus $\zeta$. On the other hand, $C_{60}$'s evolution in Fig. 4(b) is a long, multi-step addition by the transformed cages and hence, there is lesser 'surprise' near low probability region. The two fullerenes, have very different information relating to self-organization. It is trivial that for both fullerenes, there are no surprises or uncertainty around probability~1. The initial probability $p_i$ at the start of fragmentation sequence is 1 for $C_{100}$. By losing half of its population at each fragmentation stage, it is reduced to 1/128 of its initial density within 6 steps and $p_{100} \to 0$ as $ln(1/p_{100}) \to 10$. The uncertainty increases as the probability reduces for $C_{100}$. The pattern is same for $C_{60}$ but the net effect is the opposite. The relative population of $C_{60}$ increases from low probability towards higher. The graph of $-p_x(\zeta)lnp_x(\zeta)$ versus $p_x(\zeta)$ for $C_{100}$ and $C_{60}$ is similar in overall profile but significantly different in specifics and details. Shannon entropy for $C_{100}$ $\{-\sum p_{100}(\zeta)lnp_{100}(\zeta)\}$ is 0.913 and for $C_{60}$ its value $\{-\sum p_{60}(\zeta)lnp_{60}(\zeta)\}$ = 8.83. The fractal dimensions associated with the two cages are $d_I^{100} = 0.367$ and $d_I^{60} = 2.201$, respectively. The probability, the associated uncertainty and the instantaneous entropy are the measures of a dissipative dynamical system that provide individual constituent-specific details like the fragmentation pattern or the profile of the emerging structure.

## 7. Conclusions

In conclusion, the information, instantaneous $(-p_x(\zeta)lnp_x(\zeta))$ and total $(I_x)$, were obtained from four ensembles of fragmenting fullerenes. The information is based on the probabilities derived from the normalized densities of fullerenes. We chose four different ensembles that composed successively increasing number of fullerenes. Each fullerene, when subjected to the internal cage-fragmentation forces, follows the fragmentation route $C_x \to C_{x-2} +$ $C_2$. The larger cages are transformed into smaller ones and the sequence continues until the conditions for fragmentation are no longer valid. We have shown that the dissipative dynamical system of the fragmenting and transforming cages has an end-directed evolution towards $C_{60}$. It acts as the sink whose basin is the entire ensemble. Tabulated in Table 1 is the total entropy for the transformation of the entire ensemble into $C_{60}$, as the sum of $I_x$ for all cages $-\sum_\zeta \sum_x p_x(\zeta) ln p_x(\zeta) \equiv \sum I_x$. Here $\sum I_x$ can be considered as the total entropic cost of the self-organization of all of the fullerenes into $C_{60}$ with $I_{60} = -\sum_\zeta p_{60}(\zeta) ln p_{60}(\zeta)$. It is trivial that $I_{60} \ll \sum I_x$. The ratio of the total information $\sum I_x$ to $I_{60}$ of the emerging $C_{60}$s can be considered the entropic cost of $C_{60}$'s emergence, $i_{60}^\Sigma \equiv I_{total}/I_{60}$. However, it is the minimum entropic cost for the transformation of the ensemble into $C_{60}$ through the series of transformations $\sum_\zeta \sum_x C_x(\zeta) \to \sum_\zeta C_{60}(\zeta)$; incomplete or repetitive stages will increase the entropic cost. Fractal dimension $d_I^x$ for each fullerene obtained from respective information shows that $C_{60}$ has the highest fractal dimension $d_I^{60} > d_I^{x>60}$. The information determines the profiles of the phase space of fragmenting and evolving species. The fractal dimension of the constituents is introduced as a measure of the dissipative dynamical system whose transformed output is a stable fullerene cage-$C_{60}$. A similar, information-theoretic model of fragmenting single-walled carbon nanotubes has also employed fractal dimension of the fragmenting species to identify and distinguish between the linear and nonlinear dissipation mechanisms [28]. In this communication, the model is used to profile the dynamical system composed of the fragmenting and emerging fullerenes. Detailed analysis of the evolving phase space will be taken up in later studies.